\documentclass[letter,aps,prd,10.5pt,
preprintnumbers,superscriptaddress,nofootinbib,amsmath,amssymb,floatfix
]{revtex4}
\usepackage{orcidlink}

\usepackage{times}
\usepackage{amsmath}
\usepackage{amssymb}
\usepackage{graphicx}
\usepackage{subfigure}
\usepackage{booktabs}
\usepackage{rotating}
\usepackage{longtable}
\usepackage{bm}
\usepackage{float}
\usepackage{epstopdf}
\usepackage{xcolor}
\usepackage{comment}
\usepackage{graphicx}
\usepackage{subfigure}
\usepackage{hyperref}
\hypersetup{
   colorlinks   =  true,
    linkcolor    = blue,
    citecolor    = red,
     urlcolor	=magenta,     
}

\newcommand{\dd}{\mathrm{d}}
\newcommand{\cQ}{\mathcal{Q}}
\newcommand{\cR}{\mathcal{R}}
\newcommand{\cT}{\mathcal{T}}

\newcommand{\beff}{\beta_{\Theta}}
\newcommand{\lth}{\lambda_{\Theta}}
\newcommand{\safeincludegraphics}[2][]{%
	\IfFileExists{#2}{\includegraphics[#1]{#2}}{%
		\fbox{\parbox{0.88\linewidth}{\centering Figure file \texttt{#2} is not present.\\
				Compile the source package containing the \texttt{figures/} directory.}}}}

\begin{document}

\title{Constrained thermodynamics and geodesic observables of an effective
	non-commutative Kerr-like black hole}

\author{H.  Hassanabadi \orcidlink{0000-0001-7487-6898}}
\email{hassanhassanabadi@mail.fresnostate.edu}
\affiliation{Physics Department, California State University, Fresno, CA 93740, USA}
\affiliation{Department   of   Physics, Faculty of Science,   University   of   Hradec   Kr\'{a}lov\'{e},  Rokitansk\'{e}ho 62, 500   03   Hradec   Kr\'{a}lov\'{e},   Czechia}
\affiliation{Khazar University, Department of Physics and Electronics, 41 Mahsati Str, AZ1096, Baku, Azerbaijan}
\author{L. A. L\'opez \orcidlink{0000-0002-6395-6777}}
\email{lalopez@uaeh.edu.mx}
\affiliation{\'Area Acad\'emica de Matem\'aticas y F\'isica, UAEH, Carretera Pachuca--Tulancingo Km. 4.5, \\C.P. 42184, Mineral de la Reforma, Hidalgo, M\'exico}
\author{N. Bret\'on  \orcidlink{0000-0002-1237-7134}}
\email{nora.breton@cinvestav.mx}
\affiliation{Physics Department, Cinvestav, P.O. Box 14-740, Mexico City, Mexico}
\author{L. M. Nieto\footnote{Corresponding author} \orcidlink{0000-0002-2849-2647}}
\email{luismiguel.nieto.calzada@uva.es}
\affiliation{Departamento de F\'{\i}sica Te\'orica, At\'omica y Optica and Laboratory for Disruptive \\ Interdisciplinary Science (LaDIS), Universidad de Valladolid, 47011 Valladolid, Spain}
\author{S. Zare \orcidlink{0000-0003-0748-3386}}
\email{soroushzrg@gmail.com}
\affiliation{Helsinki Institute of Physics, University of Helsinki, P.O. Box 64, 00014 Helsinki, Finland}
	
		\date{\today}

	\begin{abstract}
	\medskip
		\noindent
		\textbf{Abstract}\\  
		We investigate the horizon structure, constrained thermodynamics, and geodesic properties of an effective Kerr-like black hole in a non-commutative  background. Deformation modifies the radial geometry through a mass-dependent charge-like contribution, while preserving the separability of the geodesic equations. 
		We determine the conditions for
		horizon existence, identify the extremal zero-temperature configuration,
		and analyze the stationary-limit surfaces and the ergoregion. Special
		attention is paid to the thermodynamic interpretation of the model,
		where the geometric Hawking quantities are distinguished from the
		conjugate variables associated with the constrained state space at fixed non-commutative  deformation parameter. 
		The canonical and grand-canonical
		heat capacities are derived to characterize their ensemble-dependent
		local thermal behavior. We also obtain the spherical photon region,
		equatorial light rings, and shadow boundary, showing that the deformation
		shifts the characteristic photon orbits inwards and reduces the overall size of the
		shadow. 
		Timelike circular motion is studied through the innermost
		stable circular orbit, where non-commutative  correction produces an
		inward shift of both the prograde and retrograde branches.
		Finally, invariant photon frequency shifts are obtained by treating the
		emitter's orbital direction and the photon's tangential emission
		direction as independent physical choices.
		\\
		
	\end{abstract}

	\maketitle

	\section{Introduction}
	
	Non-commutative  geometry provides a possible effective description of short-distance spacetime structure and has appeared in string theory, quantum field theory, and various approaches to quantum gravity \cite{SeibergWitten,Szabo,DouglasNekrasov}. In black hole physics, non-commutative-inspired models often replace pointlike sources by smeared distributions or introduce deformation parameters that modify the near-horizon geometry. These modifications can alter the  structure of the event horizon, Hawking evaporation, thermodynamic stability, photon motion, and observable optical signatures \cite{NicoliniReview,SmailagicSpallucci,Myung,ModestoNicolini,chaichian2008corrections,heidari2024exploring,anacleto2020absorption,anacleto2023absorption,zhao2024quasinormal,araujo2025non,heidari2025non,sharif2011thermodynamics,nozari2006reissner,banerjee2008noncommutative}.
	
	Rotating black holes are particularly relevant because astrophysical compact objects are expected to carry angular momentum. In the Kerr geometry, frame-dragging produces an ergoregion, separates prograde and retrograde photon orbits, deforms the shadow, and modifies the energetics and stability of equatorial circular motion \cite{Kerr,Carter,Chandrasekhar,Bardeen,Teo}. Therefore, it is useful to determine how an effective non-commutative  deformation changes these properties.

	Equatorial timelike circular orbits provide a direct connection between
	the geometry of a rotating black hole and the dynamics of matter in its
	strong-field region
	\cite{Carter,Bardeen1972,Pugliese2011}.
	The corresponding specific energy, specific angular momentum, and
	angular velocity determine the binding efficiency and local kinematics
	of the near-Keplerian matter surrounding the compact object. In stationary
	and axisymmetric spacetimes, frame-dragging effects separate the orbital
	motion into prograde and retrograde branches, which generally possess
	different characteristic radii and stability properties. 
	Circular
	configurations are obtained by requiring the simultaneous  cancellation of
	the effective radial  potential and its first derivative, while the
	sign of the second radial derivative determines whether the orbit is
	stable or unstable. The limiting configuration between these two
	regimes defines the innermost stable circular orbit (ISCO), whose energy and
	radius play a central role in relativistic accretion disk models
	\cite{NovikovThorne1973,PageThorne1974,Thorne1974,Cunningham1975}.
	Since  ISCO probes the spacetime geometry near the event horizon,
	it is particularly sensitive to deviations from the Kerr solution and
	can therefore serve as a useful diagnostic tool for modified, regular, or
	quantum-corrected black hole geometries
	\cite{StuchlikSlany2004,JohannsenPsaltis2011,Bambi2017,StuchlikSchee2015}.
	In the present analysis, we derive the two circular orbit branches and
	investigate how the non-commutative  deformation modifies their conserved
	quantities, orbital frequencies, and marginal stability radii. We also   explicitly
	verify that the standard Kerr expressions are recovered when
	the non-commutative  parameter vanishes.

	The effective static geometry considered here originates from a
	 Lorentz-smeared mass distribution inspired by  non-commutative
	Schwarzschild spacetime~\cite{Anacleto2020}. In the regime
	$r\gg\sqrt{\Theta}$, the corresponding lapse function takes the form \cite{araujo2024effects}
	\begin{equation}
		f(r)=1-\frac{2M}{r}
		+\frac{8M\sqrt{\Theta}}{\sqrt{\pi}\,r^{2}}
		+\mathcal{O}\!\left(
		\frac{M\Theta^{3/2}}{r^{4}}
		\right).
		\label{static_lapse}
	\end{equation}
 Next, we adopt the displayed truncated expression  shown as the
	defining lapse function of the effective model. 
	 Consequently, the results derived below are accurate for this effective metric,
while the conclusions regarding deep-core and extreme regimes should be interpreted with due caution, their rotational counterpart having a Boyer-Lindquist form similar to Kerr's. A key structural observation is that the radial metric function is algebraically identical to the  Kerr-Newman metric after the replacement
	\begin{equation}
		Q^{2}\to \beff\equiv \frac{8M\sqrt{\Theta}}{\sqrt{\pi}}.
		\label{eq:effective-beta}
	\end{equation}
	However, $\beff$ is tied to $M$ and $\Theta$ rather than being an independent electromagnetic charge. Consequently,  the familiar Kerr-Newman geodesic formulas can be adapted, but the thermodynamic interpretation requires  caution, since varying $M$ also varies $\beff$ at fixed $\Theta$.
	
	The aims of this work are fourfold. First, we determine the horizon-existence bound, the extremal zero-temperature configuration, the stationary-limit surfaces, and the ergoregion within the adopted effective geometry. Second, we formulate the constrained thermodynamics for fixed $\Theta$, explain how the entropy changes under  variations of fixed $a$, $J$, $\widetilde\Omega$, and $\Omega_H$, and separate the strict canonical and grand-canonical ensembles from the geometric Hawking responses. Third, we provide a consistent treatment of the spherical photon region, equatorial light rings, shadow, timelike circular motion, and the non-commutative  displacement of the ISCO. Fourth, we derive the frequency shifts invariantly and distinguish the emitter-orbit sign $s$ from the tangential photon-direction sign $\sigma$.
	Throughout the paper we use geometrized units $G=c=\hbar=k_{\rm B}=1$ and metric signature $(-,+,+,+)$.

	\section{Effective non-commutative  Kerr geometry}
	\label{sec:geometry}
	
The purpose of this section is to define the effective rotating geometry and to establish the domain in parameter space for which it describes a black hole. The distinction between an exact effective metric and a truncated approximation to an underlying non-commutative-inspired solution is important, particularly in the near-extremal regime where the characteristic horizon radius can become comparable to $\sqrt{\Theta}$. In the following, all results are understood as exact consequences of the metric ansatz introduced below. The geometry retains the algebraic Carter structure of the Kerr--Newman spacetime under the formal replacement $Q^2\rightarrow\beta_{\Theta}$, although $\beta_{\Theta}$ is fixed by $M$ and $\Theta$ and does not represent an independent electromagnetic charge. The construction should therefore be regarded as an effective non-commutative-inspired rotating geometry rather than as a solution derived from a complete microscopic non-commutative gravity theory. In particular, the spacetime retains the Kerr ring singularity at $\Sigma=0$.

	\subsection{Metric and useful notation}
	
	To construct the rotating counterpart of the static geometry, we employ
	the Newman--Janis algorithm (NJA), originally introduced to generate the
	Kerr metric from the Schwarzschild spacetime \cite{CapozzielloJCAP2023,CapozzielloPoDU2025}
	and subsequently extended
	to obtain the Kerr--Newman solution
	\cite{NewmanJanis1965,NewmanEtAl1965}. Since then, the method has been applied to a wide class of static seed geometries, although its implementation requires special care because the prescription for complexification is not unique \cite{DrakeSzekeres2000,BambiModesto2013,AzregAinou2014,AzregAinou2014EPJC,Erbin2017}.
	
	Consider a generic static and spherically symmetric seed metric of the form
	\begin{equation}
		ds^{2}
		=
		-F(r)\,dt^{2}
		+\frac{dr^{2}}{G(r)}
		+H(r)
		\left(
		d\theta^{2}
		+\sin^{2}\theta\,d\phi^{2}
		\right),
		\label{generic_seed_metric}
	\end{equation}
	where $F(r)$, $G(r)$, and $H(r)$ are real functions of the radial
	coordinate. We first introduce an outgoing Eddington--Finkelstein-type
	null coordinate $u$ through
	\begin{equation}
		du
		=
		dt-\frac{dr}{\sqrt{F(r)G(r)}},
		\label{null_coordinate}
	\end{equation}
	and the line element then becomes
	\begin{equation}
		ds^{2}
		=
		-F(r)\,du^{2}
		-2\sqrt{\frac{F(r)}{G(r)}}\,du\,dr
		+H(r)
		\left(
		d\theta^{2}
		+\sin^{2}\theta\,d\phi^{2}
		\right).
		\label{seed_null_coordinates}
	\end{equation}
	The corresponding inverse metric is expressed in terms of a null tetrad
	$\left(l^{\mu},n^{\mu},m^{\mu},\bar{m}^{\mu}\right)$ according to
	\begin{equation}
		g^{\mu\nu}
		=
		-l^{\mu}n^{\nu}
		-l^{\nu}n^{\mu}
		+m^{\mu}\bar{m}^{\nu}
		+m^{\nu}\bar{m}^{\mu},
		\label{null_tetrad_decomposition}
	\end{equation}
	where the tetrad vectors satisfy
	\begin{equation}
		l_{\mu}n^{\mu}=-1,
		\qquad
		m_{\mu}\bar{m}^{\mu}=1,
	\end{equation}
	with all other scalar products vanishing.
	Rotation is introduced through the complex coordinate transformation
	\begin{equation}
		r\to r'=r+ia\cos\theta,
		\qquad
		u\to u'=u-ia\cos\theta,
		\label{complex_NJA_transformation}
	\end{equation}
	where $a$ denotes the rotation parameter. Simultaneously, the radial
	functions appearing in the seed geometry are promoted to real functions
	of $r$ and $\theta$,
	\begin{equation}
		F(r)\to \widetilde{F}(r,\theta),
		\qquad
		G(r)\to \widetilde{G}(r,\theta),
		\qquad
		H(r)\to \widetilde{H}(r,\theta),
		\label{complexified_functions}
	\end{equation}
	such that
	\begin{equation}
		\lim_{a\rightarrow 0}\widetilde{F}(r,\theta)=F(r),
		\qquad
		\lim_{a\rightarrow 0}\widetilde{G}(r,\theta)=G(r),
		\qquad
		\lim_{a\rightarrow 0}\widetilde{H}(r,\theta)=H(r).
		\label{static_limit_functions}
	\end{equation}
	These conditions guarantee that the transformed metric remains real and
	reduces continuously to the original static spacetime in the
	non-rotating limit. The transformed null tetrad is then used to
	reconstruct the contravariant metric, which is subsequently inverted to
	obtain the covariant line element in null coordinates.
	
	To express the resulting stationary and axisymmetric geometry in
	Boyer--Lindquist-like coordinates, we perform a transformation of the
	form
	\begin{equation}
		du=dt+\lambda(r)\,dr,
		\qquad
		d\phi=d\varphi+\chi(r)\,dr,
		\label{BL_transformation}
	\end{equation}
	where the functions $\lambda(r)$ and $\chi(r)$ are chosen to eliminate
	the $dt\,dr$ and $dr\,d\varphi$ cross terms. Such a transformation is
	well defined only when $\lambda$ and $\chi$ depend exclusively on the
	radial coordinate; this requirement is closely related to the
	circularity of the generated rotating spacetime
	\cite{AzregAinou2014,Shaikh2019,LimaJunior2020}.

	It should be emphasized that the Newman--Janis procedure is a
	solution-generating prescription rather than a general theorem.
	Consequently, for an arbitrary seed metric, the fact that the static
	geometry satisfies a given set of field equations does not automatically
	ensure that its Newman--Janis rotating extension satisfies the same
	system. The generated metric should therefore be regarded as a candidate
	rotating geometry until it has been substituted into the complete
	Einstein--Kaluza--Klein equations, including the scalar and vector-field
	equations, and verified explicitly \cite{HansenYunes2013,LimaJunior2020}.
	\\
	The static seed metric is
	\begin{equation}
		\dd s^{2}=-f(r)\dd t^{2}+\frac{\dd r^{2}}{f(r)}
		+r^{2}\left(\dd\theta^{2}+\sin^{2}\theta\,\dd\phi^{2}\right),
	\end{equation}
	with $f(r)$ given by Eq.~\eqref{static_lapse}. The rotating counterpart is written as
	\begin{equation}
		\dd s^{2}={}-\frac{\Delta}{\Sigma}
		\left(\dd t-a\sin^{2}\theta\,\dd\phi\right)^{2}
		+\frac{\Sigma}{\Delta}\dd r^{2}+\Sigma\dd\theta^{2}
		+\frac{\sin^{2}\theta}{\Sigma}
		\left[a\dd t-(r^{2}+a^{2})\dd\phi\right]^{2},
		\label{eq:metric}
	\end{equation}
	where
	\begin{equation}
		\Sigma=r^{2}+a^{2}\cos^{2}\theta,
		\qquad
		\Delta=r^{2}-2Mr+a^{2}+\beff,
		\qquad
		\beff=\frac{8M\sqrt{\Theta}}{\sqrt{\pi}}.
		\label{eq:DeltaSigma}
	\end{equation}
	It is also convenient to define the mass-independent length
	\begin{equation}
		\lth\equiv\frac{\beff}{M}=8\sqrt{\frac{\Theta}{\pi}},
		\label{eq:lambdaTheta}
	\end{equation}
	so that $\Delta=r^{2}-2Mr+a^{2}+M\lth$.
	The metric components relevant below are
	\begin{eqnarray}
		&&g_{tt}=-\frac{\Delta-a^{2}\sin^{2}\theta}{\Sigma},\qquad\qquad
		g_{t\phi}=-\frac{a\sin^{2}\theta\,(2Mr-\beff)}{\Sigma},\\
		&&g_{\phi\phi}=\frac{\sin^{2}\theta}{\Sigma}
		\left[(r^{2}+a^{2})^{2}-a^{2}\Delta\sin^{2}\theta\right],\qquad
		g_{rr}=\frac{\Sigma}{\Delta},
		\qquad g_{\theta\theta}=\Sigma,
	\end{eqnarray}
	and they satisfy
	\begin{equation}
		g_{t\phi}^{2}-g_{tt}g_{\phi\phi}=\Delta\sin^{2}\theta.
		\label{eq:metricdetidentity}
	\end{equation}
 In the notation of that construction, the static seed is specified by
\begin{equation}
	G(r)=F(r)=f(r),
	\qquad
	H(r)=r^{2}.
\end{equation}
The auxiliary function is then
\begin{equation}
	K(r)=H(r)\sqrt{\frac{F(r)}{G(r)}}=r^{2},
\end{equation}
and the rotating functions become
\begin{equation}
	\Sigma=K(r)+a^{2}\cos^{2}\theta
	=r^{2}+a^{2}\cos^{2}\theta,
	\qquad
	\Delta=F(r)H(r)+a^{2}
	=r^{2}-2Mr+a^{2}+\beta_{\Theta}.
	\label{28j1234}
\end{equation}
Choosing the conformal function of the rotating construction as
$\Psi=\Sigma$ reproduces the Boyer--Lindquist metric in
Eq.~\eqref{eq:metric} without complexifying either the radial
coordinate or the parameters of the static seed.

	\subsection{Horizons and the bound on the non-commutative  parameter}
	
\begin{figure}[htb]
		\centering
		\safeincludegraphics[width=.45\linewidth]{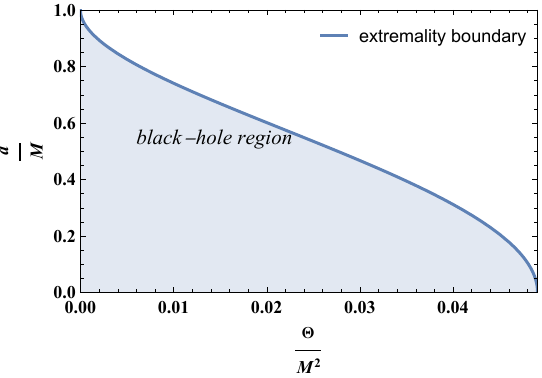}
		\caption{Allowed black hole region in the $(\Theta/M^{2},a/M)$ plane. The boundary is the extremality curve, which corresponds to the equality in Eq.~\eqref{eq:dimensionless-bound}.}
		\label{fig:parameter-region}
	\end{figure}

\begin{figure}[htb]
		\centering
		\safeincludegraphics[width=.9\linewidth]{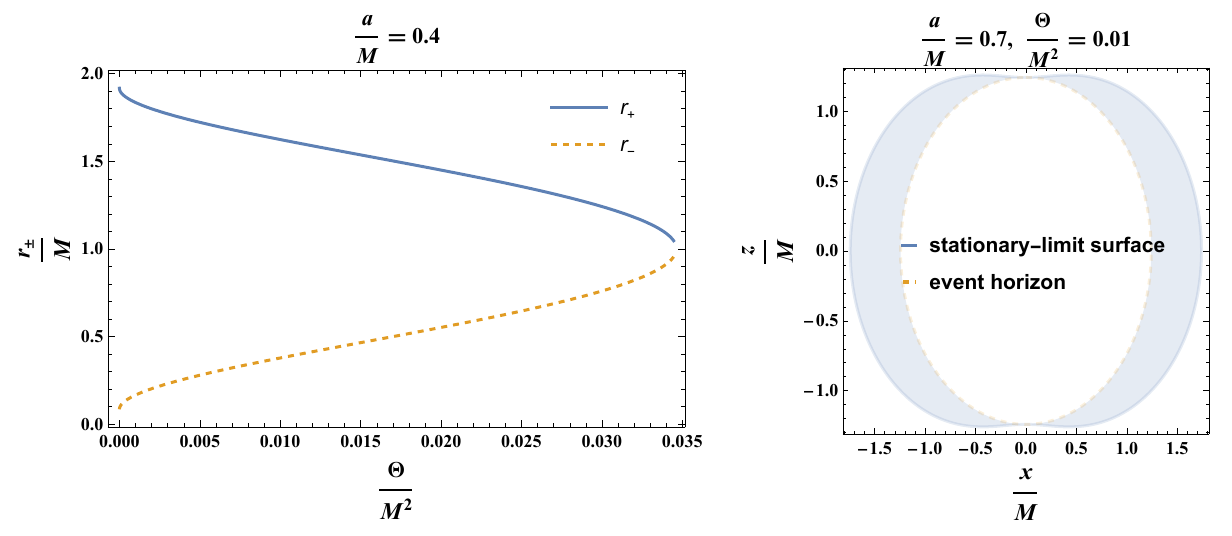}
		\caption{Left:  event horizon (blue) and Cauchy (orange) horizon as functions of $\Theta/M^{2}$ for $a/M=0.4$. Right: meridional section of the event horizon and outer stationary-limit surface. The shaded region is the ergoregion.}
		\label{fig:horizons-ergosphere}
	\end{figure}

	The Killing horizons are the roots of $\Delta=0$  in \eqref{28j1234}:
	\begin{equation}
		r_{\pm}=M\pm\sqrt{M^{2}-a^{2}-\beff}
		=M\pm\sqrt{M^{2}-a^{2}-\frac{8M\sqrt{\Theta}}{\sqrt{\pi}}}.
		\label{eq:horizons}
	\end{equation}
	The outer root $r_{+}$ is the event horizon and the inner root $r_{-}$ is the Cauchy horizon. Equivalently, for fixed $M$ and $\Theta$, the allowed spin satisfies
	$a^{2}\leq M^{2}-8M\sqrt{\Theta/\pi}$.
	Solving for $\Theta$ gives
	\begin{equation}
		\Theta\leq\Theta_{\max}
			=\frac{\pi(M^{2}-a^{2})^{2}}{64M^{2}},
		\label{eq:theta-bound}
	\end{equation}
	or, in dimensionless form,
	\begin{equation}
		\frac{\Theta}{M^{2}}\leq
		\frac{\pi}{64}\left(1-\frac{a^{2}}{M^{2}}\right)^{2}.
		\label{eq:dimensionless-bound}
	\end{equation}
	 The region described by the above inequality is represented in FIG.\ \ref{fig:parameter-region}. The blue curve (extremality boundary of the black hole region) corresponds to the equality in \eqref{eq:dimensionless-bound}.
At fixed $M$ and $a$, increasing $\Theta$ decreases the outer-horizon
radius and increases the inner-horizon radius. The separation between
the two horizons therefore shrinks until they merge at extremality as one can see the left panel of FIG.~\ref{fig:horizons-ergosphere}.
	The strict inequality corresponds to two distinct horizons, equality to a degenerate extremal horizon, and violation of the bound to a horizonless geometry. Equivalently,
	\begin{equation}
		a^{2}\leq M^{2}-\frac{8M\sqrt{\Theta}}{\sqrt{\pi}}.
	\end{equation}
	Thus, increasing $\Theta$ lowers the maximum allowed rotation parameter.
	
	\subsection{Hawking temperature and the extremal zero-temperature configuration}
	
	The surface gravity gives 
	\begin{equation}
		T_{H}=\frac{\Delta'(r_{+})}{4\pi(r_{+}^{2}+a^{2})}
		=\frac{r_{+}-M}{2\pi(r_{+}^{2}+a^{2})}.
		\label{eq:temperature-basic}
	\end{equation}
	Using $\Delta(r_{+})=0$, the mass can be parametrized by the horizon radius as
	\begin{equation}
		M(r_{+})=\frac{r_{+}^{2}+a^{2}}{2r_{+}-\lth}
		=\frac{\sqrt{\pi}(r_{+}^{2}+a^{2})}
		{2(\sqrt{\pi}r_{+}-4\sqrt{\Theta})}.
		\label{eq:mass-rh}
	\end{equation}
	The temperature then takes the transparent form
	\begin{equation}
		T_{H}(r_{+})=
			\frac{r_{+}^{2}-\lth r_{+}-a^{2}}
			{2\pi(r_{+}^{2}+a^{2})(2r_{+}-\lth)}.
		\label{eq:temperature-rh}
	\end{equation}
	The zero-temperature endpoint is determined by
	\begin{equation}
		r_{\rm rem}^{2}-\lth r_{\rm rem}-a^{2}=0,
	\end{equation}
	so the positive solution is
	\begin{equation}
		r_{\rm rem}=
			\frac{\lth+\sqrt{\lth^{2}+4a^{2}}}{2}
			=\frac{4\sqrt{\Theta}}{\sqrt{\pi}}
			+\sqrt{a^{2}+\frac{16\Theta}{\pi}}.
		\label{eq:remnant-radius}
	\end{equation}
	At extremality $\Delta=\Delta'=0$, and since $\Delta'=2(r-M)$,  we have that
	$M_{\rm rem}=r_{\rm rem}$.
	The Kerr limit gives $M_{\rm rem}=r_{\rm rem}=|a|$. In the non-rotating limit,
	\begin{equation}
		M_{\rm rem}=r_{\rm rem}=\frac{8\sqrt{\Theta}}{\sqrt{\pi}}.
	\end{equation}
	 In the simultaneous limits $a\to0$ and $\Theta\to0$, the extremal
	length scale shrinks to zero, consistently with the absence of a
	finite zero-temperature endpoint for the Schwarzschild black hole.

	\subsection{Stationary-limit surfaces and the ergoregion}
	\label{subsec:ergosphere}
	
	The stationary-limit surfaces are defined by $g_{tt}=0$, not by $\Delta=0$. From Eq.~\eqref{eq:metric},
	\begin{equation}
		g_{tt}=0
		\quad\Longleftrightarrow\quad
		r^{2}-2Mr+a^{2}\cos^{2}\theta+\beff=0.
	\end{equation}
	Therefore,
	\begin{equation}
		r_{\rm e}^{\pm}(\theta)=M\pm
			\sqrt{M^{2}-a^{2}\cos^{2}\theta-\beff}.
		\label{eq:stationary-surfaces}
	\end{equation}
	The physically relevant outer surface is $r_{\rm e}^{+}(\theta)$, and the ergoregion is
	\begin{equation}
		r_{+}<r<r_{\rm e}^{+}(\theta).
		\label{eq:ergoregion}
	\end{equation}
	Within this region the asymptotic time-translation Killing vector
	$\partial_t$ becomes spacelike. Consequently, no physical observer can
	remain at fixed spatial Boyer--Lindquist coordinates relative to
	asymptotic infinity.
	
	At the poles, $\cos^{2}\theta=1$, and the stationary-limit surface touches the event horizon:
	$r_{\rm e}^{+}(0)=r_{+}$.
	At the equator,
	\begin{equation}
		r_{\rm e}^{+}\!\left(\frac{\pi}{2}\right)
		=M+\sqrt{M^{2}-\beff},
	\end{equation}
	and the equatorial thickness of the ergoregion is
	\begin{equation}
		\Delta r_{\rm ergo}^{\rm(eq)}=
		\sqrt{M^{2}-\beff}-\sqrt{M^{2}-a^{2}-\beff}.
	\end{equation}
Although both the outer horizon and the equatorial stationary-limit
surface move inward as $\Theta$ increases, their difference grows at
fixed $M$ and $a$. Thus, the non-commutative  deformation increases the
equatorial thickness of the ergoregion while reducing its overall
radial scale.
	For $a=0$, the stationary-limit and horizon surfaces coincide, so there is no genuine ergoregion. In the extremal case $M^{2}-a^{2}-\beff=0$,
	\begin{equation}
		r_{\rm e}^{+}(\theta)=M+|a|\sin\theta,
		\qquad r_{+}=M.
	\end{equation}
 All these regions and curves are shown in FIG.~\ref{fig:horizons-ergosphere}.

	\subsection{Small-$\Theta$ expansions}
	
	Because the metric correction is proportional to $\sqrt{\Theta}$, the natural perturbative parameter is $\beff=\mathcal{O}(\sqrt{\Theta})$. For the horizons, define $A=M^{2}-a^{2}$ and assume $\beff/A\ll1$. Then
	\begin{align}
		r_{\pm}\simeq{}&M\pm\sqrt{A}
		\mp\frac{4M\sqrt{\Theta}}{\sqrt{\pi A}}
		\mp\frac{8M^{2}\Theta}{\pi A^{3/2}}
		+\mathcal{O}(\Theta^{3/2}).
		\label{eq:horizon-expansion}
	\end{align}
	For the stationary-limit surfaces define $A_{\theta}=M^{2}-a^{2}\cos^{2}\theta$. Their second-order expansion is
	\begin{align}
		r_{\rm e}^{\pm}(\theta)\simeq{}&M\pm\sqrt{A_{\theta}}
		\mp\frac{4M\sqrt{\Theta}}{\sqrt{\pi A_{\theta}}}
		\mp\frac{8M^{2}\Theta}{\pi A_{\theta}^{3/2}}
		+\mathcal{O}(\Theta^{3/2}).
		\label{eq:stationary-expansion}
	\end{align}
	These expansions are not reliable close to extremality, where the denominator becomes comparable to the deformation term.

	\section{CONSTRAINED THERMODYNAMICS AND LOCAL STABILITY}
	\label{sec:thermo}
	
The thermodynamic interpretation of the present family requires
special care. The surface-gravity temperature $T_H$ and the horizon
angular velocity $\Omega_H$ are geometric quantities associated with
the Killing horizon. However, the Kerr--Newman-like deformation
parameter is constrained by $\beta_{\Theta}=M\lambda_{\Theta}$ and
therefore changes when the mass is varied at fixed $\Theta$. As a
result, the usual Kerr first law cannot be imported without accounting
for the induced variation of the charge-like term. We shall
distinguish the geometric horizon quantities from the reduced
thermodynamic conjugates obtained after restricting the state space to
the constraint $\beta_{\Theta}=M\lambda_{\Theta}$. Because the geometry is asymptotically flat, the response functions
considered below diagnose local stability within the reduced state
space. A globally stable canonical or grand-canonical ensemble would
require an additional confining prescription, such as a finite
thermodynamic cavity.

	\subsection{Geometric horizon quantities and constrained thermodynamic conjugates}
	
	For compactness in this section we write $r\equiv r_{+}$ and define
	\begin{equation}
		\mathcal{K}=r^{2}+a^{2},
		\qquad
		N=r^{2}-a^{2}-\lth r.
		\label{eq:thermo-KN}
	\end{equation}
	The horizon equation \eqref{eq:mass-rh} gives
	\begin{equation}
		M=\frac{\mathcal{K}}{2r-\lth},
		\qquad
		S=\pi\mathcal{K},
		\qquad
		J=aM.
		\label{eq:MSJ-parametric}
	\end{equation}

\subsubsection{Horizon entropy and its dependence on the deformation
	parameter}

Throughout this section, we assume that the gravitational sector is
described by the Einstein--Hilbert action and that the
non-commutative  correction is encoded in an effective matter source.
Under this assumption, the horizon entropy obeys the
Bekenstein--Hawking area law.

At $r=r_{+}$, the nonvanishing components of the induced horizon
two-metric are
\begin{equation}
	g_{\theta\theta}=\Sigma,
	\qquad
	g_{\phi\phi}
	=\frac{(r^{2}+a^{2})^{2}\sin^{2}\theta}{\Sigma}.
\end{equation}
Consequently,
\begin{equation}
	A_H
	=\int_{0}^{2\pi}d\phi\int_{0}^{\pi}d\theta\,
	\sqrt{g_{\theta\theta}g_{\phi\phi}}
	=4\pi(r^{2}+a^{2}),
\end{equation}
and
\begin{equation}
	S=\frac{A_H}{4}
	=\pi(r^{2}+a^{2}).
\end{equation}
	The local area density is proportional to $\sin\theta$, but the integration over the full horizon removes the polar-coordinate dependence. The local area density depends on the polar coordinate through
	$\sin\theta$, but the integrated horizon area and entropy are
	independent of $\theta$.
	Moreover, $S(r,a)$ contains no explicit occurrence of $\Theta$,
	\begin{equation}
		\left(\frac{\partial S}{\partial\Theta}\right)_{r,a}=0,
		\label{eq:entropy-explicitTheta}
	\end{equation}
	but it depends implicitly on $\Theta$ whenever the horizon position or the rotation parameter changes with the deformation. A derivative with respect to $\Theta$ is therefore meaningful only after specifying which additional state variables are held fixed.
	
	For fixed $(M,a)$, the outer horizon is
	\begin{equation}
		r=M+\sqrt{M^{2}-a^{2}-M\lth},
	\end{equation}
	so that
	\begin{equation}
			\left(\frac{\partial S}{\partial\Theta}\right)_{M,a}
			=-\frac{4\sqrt{\pi}\,Mr}
			{\sqrt{\Theta}\,(r-M)}<0.
		\label{eq:dS-dTheta-Ma}
	\end{equation}
	At fixed $(M,J)$ one has $a=J/M={\rm const}$.
Since fixing both $M$ and $J$ also fixes $a=J/M$, the same derivative
applies to the fixed-$(M,J)$ variation.
	Thus, at fixed mass and angular momentum, increasing the non-commutative  parameter decreases the outer-horizon radius, area, and entropy.
	
	For fixed $(M,\Omega_H)$, both $r$ and $a$ vary while
	$\Omega_H=a/(r^{2}+a^{2})$ remains constant. Defining
	\begin{equation}
		\mathcal H_{\Omega_H}
		=r(r^{2}+a^{2})-M(r^{2}-a^{2}),
	\end{equation}
	one obtains
	\begin{equation}
			\left(\frac{\partial S}{\partial\Theta}\right)_{M,\Omega_H}
			=-\frac{4\sqrt{\pi}\,Mr(r^{2}+a^{2})}
			{\sqrt{\Theta}\,\mathcal H_{\Omega_H}}.
		\label{eq:dS-dTheta-MOmegaH}
	\end{equation}
	For the physical outer-horizon branch $\mathcal H_{\Omega_H}>0$, so the entropy again decreases as $\Theta$ increases.
	
	At fixed $\Theta$, the entropy derivatives along the state-space paths used below are
	\begin{align}
		\left(\frac{\dd S}{\dd r}\right)_{a,\Theta}
		&=2\pi r,
		\label{eq:dSdr-fixeda}\\
		\left(\frac{\dd S}{\dd r}\right)_{J,\Theta}
		&=\frac{2\pi(r^{2}+a^{2})[2(r^{2}+a^{2})-\lth r]}
		{(2r-\lth)(r^{2}+3a^{2})},
		\label{eq:dSdr-fixedJ}\\
		\left(\frac{\dd S}{\dd r}\right)_{\widetilde\Omega,\Theta}
		&=\frac{2\pi(r^{2}+a^{2})(2r-\lth)}
		{2(r^{2}-a^{2})-\lth r},
		\label{eq:dSdr-fixedOtilde}\\
		\left(\frac{\dd S}{\dd r}\right)_{\Omega_H,\Theta}
		&=\frac{2\pi r(r^{2}+a^{2})}{r^{2}-a^{2}}.
		\label{eq:dSdr-fixedOmegaH}
	\end{align}
	These expressions make explicit that ``fixed $a$'', ``fixed $J$'', ``fixed $\widetilde\Omega$'', and ``fixed $\Omega_H$'' correspond to different paths through the thermodynamic state space.
	
	The geometric Hawking temperature and horizon angular velocity are
	\begin{equation}
		T_{H}=\frac{N}{2\pi\mathcal{K}(2r-\lth)},
		\qquad
		\Omega_{H}=\frac{a}{\mathcal{K}}.
		\label{eq:geometric-TH-OH}
	\end{equation}
	These are determined directly by the surface gravity and the horizon generator. However, the deformation obeys $\beff=M\lth$; therefore, at fixed $\Theta$ the effective Kerr--Newman-like term changes when $M$ changes. The standard relation $\dd M=T_H\dd S+\Omega_H\dd J$ is consequently not the exact differential first law of this constrained family.
	
	Eliminating $r$ and $a$ from Eq.~\eqref{eq:MSJ-parametric}, with $\mathcal{K}=S/\pi$, gives
	\begin{equation}
		(4\mathcal{K}-\lth^{2})M^{2}
		-2\lth\mathcal{K}M-\mathcal{K}^{2}-4J^{2}=0.
		\label{eq:fundamental-quadratic}
	\end{equation}
	The positive mass branch is
	\begin{equation}
		M(S,J,\Theta)=
			\frac{\lth\mathcal{K}
				+2\sqrt{\mathcal{K}^{3}+J^{2}(4\mathcal{K}-\lth^{2})}}
			{4\mathcal{K}-\lth^{2}},
		\qquad \mathcal{K}=\frac{S}{\pi}.
		\label{eq:fundamental-mass}
	\end{equation}
	At fixed $\Theta$ its exact differential is
	\begin{equation}
		\dd M=\widetilde T\,\dd S+\widetilde\Omega\,\dd J,
		\label{eq:constrained-first-law}
	\end{equation}
	where
	\begin{eqnarray}
		&&\widetilde T
			=\left(\frac{\partial M}{\partial S}\right)_{J,\Theta}
			=\frac{N}{\pi(2r-\lth)(2\mathcal{K}-\lth r)},
		\label{eq:Ttilde}\\
		&&\widetilde\Omega
			=\left(\frac{\partial M}{\partial J}\right)_{S,\Theta}
			=\frac{2a}{2\mathcal{K}-\lth r}.
		\label{eq:Otilde}
	\end{eqnarray}
	Equivalently,
	\begin{equation}
		\widetilde T=\frac{T_H}{\Gamma_\Theta},
		\qquad
		\widetilde\Omega=\frac{\Omega_H}{\Gamma_\Theta},
		\qquad
		\Gamma_\Theta=1-\frac{\lth r}{2\mathcal{K}}.
		\label{eq:Gamma-factor}
	\end{equation}
	Equivalently, the geometric quantities satisfy the constrained
	relation
	\begin{equation}
		\Gamma_\Theta\,dM
		=
		T_H\,dS+\Omega_H\,dJ.
	\end{equation}
	The Hawking temperature measured through the radiation spectrum
	remains $T_H$; the quantity $\widetilde T$ is the conjugate temperature
	of the reduced fundamental relation.
	Thus $T_H$ and $\Omega_H$ remain the geometric horizon quantities, Thus, $T_H$ and $\Omega_H$ remain the geometric Hawking temperature
	and horizon angular velocity, whereas $\widetilde T$ and
	$\widetilde\Omega$ are the reduced thermodynamic conjugates obtained
	after restricting the state space to
	$\beta_\Theta=M\lambda_\Theta$ at fixed $\Theta$. All four coincide pairwise in the Kerr limit $\Theta\to0$. If $\Theta$ is varied, an additional work term $\Psi_\Theta\dd\Theta$ must be included.
	
	For reference, the on-shell geometric combination is
	\begin{equation}
		F_H\equiv M-T_HS
		=\frac{r^{2}+3a^{2}+\lth r}{2(2r-\lth)}.
		\label{eq:FH-geometric}
	\end{equation}
	At $\Theta\neq0$, $F_H$ should not be identified with the strict fixed-$J$ Helmholtz potential because $T_H\neq\widetilde T$.

	\subsection{Geometric Hawking response at fixed rotation parameter $a$}
	\label{subsec:fixeda}
	
	Holding $a$ and $\Theta$ fixed is a useful geometric variation, but it is not the standard canonical ensemble because $J=aM$ changes when the mass changes. Equivalently, the geometric quantities satisfy the constrained
	relation
	\begin{equation}
		\Gamma_\Theta\,dM
		=
		T_H\,dS+\Omega_H\,dJ.
	\end{equation}
	The Hawking temperature measured through the radiation spectrum
	remains $T_H$; the quantity $\widetilde T$ is the conjugate temperature
	of the reduced fundamental relation.
	and the Hawking-temperature response is
	\begin{equation}
		C^{(H)}_{a,\Theta}
		=
		T_H\left(\frac{\partial S}{\partial T_H}\right)_{a,\Theta}
		=
		\frac{
			\pi r(r^2+a^2)(2r-\lambda_\Theta)N
		}{
			D_a
		},
		\label{fixed_a_heat_capacity}
	\end{equation}
	where
	\begin{equation}
			\mathcal D_a=
			a^{4}+4a^{2}r^{2}-r^{4}
			+\frac{\lth^{2}}{2}(a^{2}-r^{2})
			-2\lth r(a^{2}-r^{2}).
		\label{eq:Da}
	\end{equation}
	The fixed-$a$ geometric critical points satisfy
	\begin{equation}
		\mathcal D_a(r_c)=0
			\quad\Longleftrightarrow\quad
			\left(\frac{\partial T_H}{\partial r}\right)_{a,\Theta}=0.
		\label{eq:critical-fixeda}
	\end{equation}
	At the extremal remnant $N=0$, both $T_H$ and $C_{a,\Theta}^{(H)}$ vanish. A divergence of $C_{a,\Theta}^{(H)}$ marks a Davies-type turning point of the geometric Hawking temperature along the fixed-$a$ path; it must not be confused with the strict fixed-$J$ canonical critical point.

	\subsection{Strict canonical ensemble at fixed $J$}
	\label{subsec:fixedJ}
	
	The canonical ensemble fixes $J$ and $\Theta$. The entropy still has the area form $S=\pi(r^{2}+a^{2})$, but $a$ is not constant along this path. Indeed, Eq.~\eqref{eq:dSdr-fixedJ} gives the corresponding entropy variation. Because
	\begin{equation}
		J=\frac{a(r^{2}+a^{2})}{2r-\lth},
	\end{equation}
	holding $J$ fixed requires $a$ to vary with $r$ according to
	\begin{equation}
		\left(\frac{\dd a}{\dd r}\right)_{J,\Theta}
			=-\frac{2a(r^{2}-a^{2}-\lth r)}
			{(2r-\lth)(r^{2}+3a^{2})}.
		\label{eq:da-dr-fixedJ}
	\end{equation}
	The strict Helmholtz potential is
	\begin{equation}
		\widetilde F_J=M-\widetilde T S,
	\end{equation}
	which becomes
	\begin{equation}
		\widetilde F_J=
			\frac{(r^{2}+a^{2})(r^{2}+3a^{2})}
			{(2r-\lth)\,[2(r^{2}+a^{2})-\lth r]}.
		\label{eq:FtildeJ}
	\end{equation}
	Its differential at fixed $\Theta$ is
	\begin{equation}
		\dd\widetilde F_J=-S\,\dd\widetilde T
		+\widetilde\Omega\,\dd J.
	\end{equation}
	The canonical heat capacity is
	\begin{equation}
		\widetilde C_{J,\Theta}
		=\widetilde T\left(\frac{\partial S}{\partial\widetilde T}\right)_{J,\Theta}
		=\widetilde C_{J,\Theta} 
		=
			\frac{2\pi N\,[2(r^{2}+a^{2})-\lth r]^{2}}
			{\widetilde{\mathcal D}_{J}},
		\label{eq:CtildeJ}
	\end{equation}
	with
	\begin{equation}
		\widetilde{\mathcal D}_{J}={}
		12a^{4}+24a^{2}r^{2}-4r^{4}
		+9a^{2}\lth^{2}-24a^{2}\lth r -3\lth^{2}r^{2}+8\lth r^{3}.
		\label{eq:DtildeJ}
	\end{equation}
	The canonical critical points satisfy
	\begin{equation}
		\widetilde{\mathcal D}_{J}(r_c)=0
			\quad\Longleftrightarrow\quad
			\left(\frac{\partial\widetilde T}{\partial r}\right)_{J,\Theta}=0.
		\label{eq:critical-fixedJ}
	\end{equation}
	At such a point $\widetilde C_{J,\Theta}$ diverges and changes sign, At this point $\widetilde C_{J,\Theta}$ diverges and changes sign,
	identifying a Davies-type critical point that separates branches with
	different local thermal responses.
	\begin{equation}
		\widetilde T(r_{\rm rem})=0,
		\qquad
		\widetilde C_{J,\Theta}(r_{\rm rem})=0.
	\end{equation}
	For the physical branches studied here, the interval immediately above the remnant has $\widetilde C_{J,\Theta}>0$, whereas the sufficiently large-radius branch has $\widetilde C_{J,\Theta}<0$.
	
	In the Kerr limit,
	\begin{equation}
		\widetilde C_J=
		\frac{2\pi(r^{2}+a^{2})^{2}(r^{2}-a^{2})}
		{3a^{4}+6a^{2}r^{2}-r^{4}},
	\end{equation}
	and the critical radius is
	\begin{equation}
		r_c=|a|\sqrt{3+2\sqrt{3}}.
	\end{equation}
	For $J=0$, hence $a=0$, the physical critical radius is
	\begin{equation}
		r_c=\frac{3}{2}\lth,
		\qquad
		r_{\rm rem}=\lth,
	\end{equation}
	so the strict canonical stable interval is $\lth<r<3\lth/2$.

	\subsection{Strict grand-canonical ensemble at fixed $\widetilde\Omega$}
	\label{subsec:grandcanonical}
	
	The thermodynamically consistent grand-canonical ensemble fixes $\widetilde T$, $\widetilde\Omega$, and $\Theta$. Along a curve of fixed $\widetilde\Omega$ and $\Theta$, $a$ varies with $r$, and the entropy derivative is given by Eq.~\eqref{eq:dSdr-fixedOtilde}. Its grand potential is
	\begin{equation}
		\widetilde G=M-\widetilde T S-\widetilde\Omega J,
	\end{equation}
	which reduces to
	\begin{equation}
		\widetilde G=
			\frac{(r^{2}+a^{2})^{2}}
			{(2r-\lth)\,[2(r^{2}+a^{2})-\lth r]}.
		\label{eq:Gtilde}
	\end{equation}
	At fixed $\Theta$,
	\begin{equation}
		\dd\widetilde G=-S\,\dd\widetilde T-J\,\dd\widetilde\Omega.
	\end{equation}
	The heat capacity at fixed $\widetilde\Omega$ is
	\begin{equation}
		\widetilde C_{\widetilde\Omega,\Theta}
			=-\frac{2\pi(2r-\lth)^{2}N}
			{4(r^{2}+a^{2})-8\lth r+3\lth^{2}}.
		\label{eq:CtildeOmega}
	\end{equation}
	Its divergences satisfy
	\begin{equation}
		4(r_c^{2}+a_c^{2})-8\lth r_c+3\lth^{2}=0.
		\label{eq:critical-tildeOmega}
	\end{equation}
	The Kerr limit is
	\begin{equation}
		\widetilde C_{\widetilde\Omega}^{\rm Kerr}
		=-\frac{2\pi r^{2}(r^{2}-a^{2})}{r^{2}+a^{2}}<0
	\end{equation}
	for every nonextremal black hole. In the static non-commutative  case, the physical critical point is again $r_c=3\lth/2$, and the interval $\lth<r<3\lth/2$ has positive heat capacity. A complete grand-canonical stability analysis also requires the rotational susceptibility
	\begin{equation}
		\chi_{\widetilde T}=\left(\frac{\partial J}{\partial\widetilde\Omega}\right)_{\widetilde T,\Theta}>0;
	\end{equation}
	positive heat capacity alone is necessary but not sufficient.

	\subsection{Hawking response at fixed geometric $\Omega_H$}
	\label{subsec:fixedOH}
	
	One may also ask how the geometric Hawking temperature changes along a curve of fixed horizon angular velocity $\Omega_H=a/(r^{2}+a^{2})$. This is distinct from fixing the exact thermodynamic conjugate $\widetilde\Omega$. The entropy variation along the fixed-$\Omega_H$ path is given by Eq.~\eqref{eq:dSdr-fixedOmegaH}. The constraint is
	\begin{equation}
		\left(\frac{\dd a}{\dd r}\right)_{\Omega_H,\Theta}
			=\frac{2ar}{r^{2}-a^{2}}.
		\label{eq:da-dr-fixedOH}
	\end{equation}
	The corresponding on-shell Gibbs-like combination is
	\begin{equation}
		G_H=M-T_HS-\Omega_HJ
		=\frac{r^{2}+a^{2}+\lth r}{2(2r-\lth)}.
		\label{eq:GH-geometric}
	\end{equation}
	The Hawking response at fixed geometric $\Omega_H$ is
	\begin{equation}
		C_{\Omega_H,\Theta}^{(H)}
			=T_H\left(\frac{\partial S}{\partial T_H}\right)_{\Omega_H,\Theta}
			=-\frac{2\pi r(2r-\lth)N}
			{2(r^{2}+a^{2})-4\lth r+\lth^{2}}.
		\label{eq:COmegaH-H}
	\end{equation}
	Its turning points obey
	\begin{equation}
		2(r_c^{2}+a_c^{2})-4\lth r_c+\lth^{2}=0,
		\qquad
		\Omega_H=\frac{a_c}{r_c^{2}+a_c^{2}}.
		\label{eq:critical-OmegaH-H}
	\end{equation}
	This quantity is useful as a geometric response function, but it is not the strict grand-canonical heat capacity because $\Omega_H\neq\widetilde\Omega$ and $T_H\neq\widetilde T$ when $\Theta\neq0$.

	\subsection{Distinct critical points and thermodynamic topology}
	
	The following conditions describe different physical variations and must not be mixed. Their entropy derivatives are likewise different, as summarized in Eqs.~\eqref{eq:dSdr-fixeda}--\eqref{eq:dSdr-fixedOmegaH}:
	\begin{align}
		\left(\frac{\partial T_H}{\partial r}\right)_{a,\Theta}=0
		&\quad\text{fixed-$(a,\Theta)$ geometric response},\\
		\left(\frac{\partial\widetilde T}{\partial r}\right)_{J,\Theta}=0
		&\quad\text{strict fixed-$J$ canonical ensemble},\\
		\left(\frac{\partial\widetilde T}{\partial r}\right)_{\widetilde\Omega,\Theta}=0
		&\quad\text{strict grand-canonical ensemble},\\
		\left(\frac{\partial T_H}{\partial r}\right)_{\Omega_H,\Theta}=0
		&\quad\text{fixed-geometric-$\Omega_H$ Hawking response}.
	\end{align}
	They generally give different critical radii at nonzero $\Theta$. In particular, a thermodynamic-topology construction must be based on the potential and conjugate temperature appropriate to the selected ensemble. An off-shell turning point obtained from $M-S/\tau$ at fixed $a$ cannot be identified with the canonical fixed-$J$ critical point or with the strict grand-canonical critical point.

	\section{Null geodesics, photon region, and shadow}
	\label{sec:photons}
	
The rotating metric preserves the Hamilton--Jacobi separability
characteristic of the Carter class. This permits the null geodesic
equations to be expressed in terms of the conserved energy,
azimuthal angular momentum, and Carter constant. In contrast with a
static spherical spacetime, where the relevant trapped null set is a
single photon sphere, a rotating black hole possesses a
two-dimensional photon region composed of spherical null trajectories
with different constant radii. The unstable subset of this region
determines the boundary of the observed black hole shadow \cite{PerlickTsupko,ZarePLB2024,ZareJCAP2024,ZareJCAP2026,ZareEPJC2026,Hassanabadi2026-1,Hosseinifar2026,Hassanabadi2026-2}.

	\subsection{Separated null-geodesic equations}
	
	Let $E=-p_{t}$, $L_{z}=p_{\phi}$, and $\cQ$ denote the photon energy, azimuthal angular momentum, and Carter constant. Define
	\begin{equation}
		P(r)=E(r^{2}+a^{2})-aL_{z}.
	\end{equation}
	The first-order null-geodesic equations are
	\begin{align}
		\Sigma\dot t&=a(L_{z}-aE\sin^{2}\theta)
		+\frac{r^{2}+a^{2}}{\Delta}P(r),\\
		\Sigma\dot\phi&=L_{z}\csc^{2}\theta-aE
		+\frac{a}{\Delta}P(r),\\
		\Sigma^{2}\dot r^{2}&=\cR(r),\\
		\Sigma^{2}\dot\theta^{2}&=\cT(\theta),
		\label{eq:null-firstorder}
	\end{align}
	The physically allowed radial and polar motions require
	$\mathcal{R}(r)\geq 0$ and $\mathcal{T}(\theta)\geq 0$, respectively with
	\begin{align}
		\cR(r)&=P(r)^{2}-\Delta\left[(L_{z}-aE)^{2}+\cQ\right],
		\label{eq:radial-potential-null}\\
		\cT(\theta)&=\cQ+\cos^{2}\theta
		\left(a^{2}E^{2}-L_{z}^{2}\csc^{2}\theta\right).
		\label{eq:angular-potential-null}
	\end{align}
	The overdot denotes differentiation with respect to an affine parameter. Introduce the impact parameters
	\begin{equation}
		\xi=\frac{L_{z}}{E},
		\qquad
		\eta=\frac{\cQ}{E^{2}}.
	\end{equation}
	Then
	\begin{equation}
		\frac{\cR}{E^{2}}=\left[(r^{2}+a^{2})-a\xi\right]^{2}
		-\Delta\left[(\xi-a)^{2}+\eta\right],\qquad
		\frac{\cT}{E^{2}}=\eta+a^{2}\cos^{2}\theta-\xi^{2}\cot^{2}\theta.
	\end{equation}

	\subsection{Spherical photon orbits and the photon region}
	
	A spherical null orbit has constant Boyer--Lindquist radius $r=r_{p}$ while it may oscillate in $\theta$. It satisfies
	\begin{equation}
		\cR(r_{p})=0,
		\qquad
		\cR'(r_{p})=0.
		\label{eq:spherical-conditions}
	\end{equation}
	Solving for the impact parameters gives
	\begin{align}
		\xi_{p}(r)&=\frac{(r^{2}+a^{2})\Delta'-4r\Delta}{a\Delta'},
		\label{eq:xi-general}\\
		\eta_{p}(r)&=\frac{16r^{2}\Delta}{(\Delta')^{2}}
		-\left(\frac{r^{2}\Delta'-4r\Delta}{a\Delta'}\right)^{2}.
		\label{eq:eta-general}
	\end{align}
	Since $\Delta'=2(r-M)$,
	\begin{eqnarray}
		&&\xi_{p}(r)=
			\frac{a^{2}(M+r)+r(r^{2}-3Mr+2\beff)}{a(M-r)},
		\label{eq:xi-expanded}
		\\
		&&\eta_p(r)=
		\frac{r^2\left[
			4a^2(Mr-\beta_\Theta)
			-\left(r^2-3Mr+2\beta_\Theta\right)^2
			\right]}
		{a^2(r-M)^2}.
		\label{eq:eta-expanded}
	\end{eqnarray}
	The polar motion is allowed when
	\begin{equation}
		\eta_{p}+a^{2}\cos^{2}\theta-\xi_{p}^{2}\cot^{2}\theta\geq0.
	\end{equation}
	Equivalently, the photon region in the $(r,\theta)$ plane obeys
	\begin{equation}
		\left(4r\Delta-\Sigma\Delta'\right)^{2}
			\leq16a^{2}r^{2}\Delta\sin^{2}\theta,
		\qquad r\geq r_{+},
		\label{eq:photon-region}
	\end{equation}
	together with the instability condition $\cR''(r_{p})>0$ for the branch that forms the shadow boundary. In a static spherical spacetime, this region collapses to a single
	photon sphere, whereas in a rotating spacetime it consists of a finite
	family of spherical photon orbits. 
	  A graph showing the corresponding shadow boundaries will be shown later.

	\subsection{Equatorial light rings and special limits}
	
\begin{figure}[htb]
		\centering
		\safeincludegraphics[width=.45\linewidth]{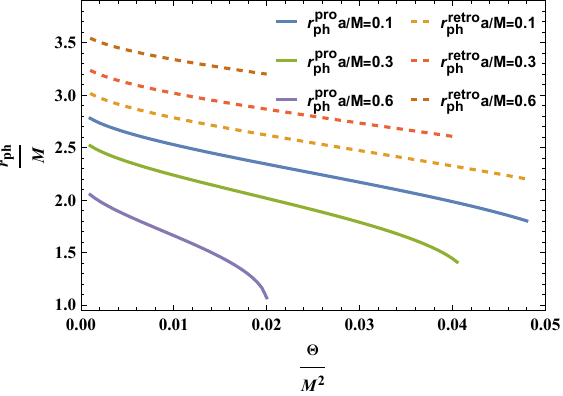}
		\caption{Prograde and retrograde equatorial photon radii as functions of the non-commutative  parameter. Each curve terminates before the corresponding extremality bound.}
		\label{fig:photon-radii}
	\end{figure}

	A photon remains in the equatorial plane when $\theta=\pi/2$ and $\cQ=0$, hence $\eta_{p}=0$. The prograde and retrograde equatorial light-ring radii satisfy
	\begin{equation}
		(r_{\rm ph}^{2}-3Mr_{\rm ph}+2\beff)^{2}
			=4a^{2}(Mr_{\rm ph}-\beff),
		\label{eq:photon-radius-squared}
	\end{equation}
	or
	\begin{equation}
		r_{\rm ph}^{2}-3Mr_{\rm ph}+2\beff
			\pm2a\sqrt{Mr_{\rm ph}-\beff}=0.
		\label{eq:photon-radius}
	\end{equation}
	The dependence of the prograde and retrograde equatorial photon radii
	on the non-commutative  parameter is shown in FIG.~\ref{fig:photon-radii}. For $a>0$, the plus-sign branch gives the smaller exterior prograde light ring and the minus-sign branch the larger retrograde light ring.
	
	For $a=0$, the outer unstable photon sphere is
	\begin{equation}
		r_{\rm ph}=\frac{1}{2}\left[3M+
			\sqrt{9M^{2}-8\beff}\right]
		=\frac{1}{2}\left[3M+
		\sqrt{9M^{2}-\frac{64M\sqrt{\Theta}}{\sqrt{\pi}}}\right].
		\label{eq:static-photon-radius}
	\end{equation}
	For small $\Theta$,
	\begin{equation}
		r_{\rm ph}=3M-\frac{16\sqrt{\Theta}}{3\sqrt{\pi}}
		+\mathcal{O}(\Theta).
	\end{equation}
	The Schwarzschild value $r_{\rm ph}=3M$ follows when $a=\Theta=0$.
	
	For $\Theta=0$, Eq.~\eqref{eq:photon-radius} reduces to the Kerr light-ring equation. In the extremal rotating case, the prograde light ring approaches the degenerate horizon, $r_{\rm ph}^{\rm pro}=r_{+}=M$. For the extremal static remnant, the outer unstable photon sphere is $r_{\rm ph}=2M_{\rm rem}$.

	\subsection{Shadow coordinates}
	
\begin{figure}[htb]
		\centering
		\safeincludegraphics[width=.75\linewidth]{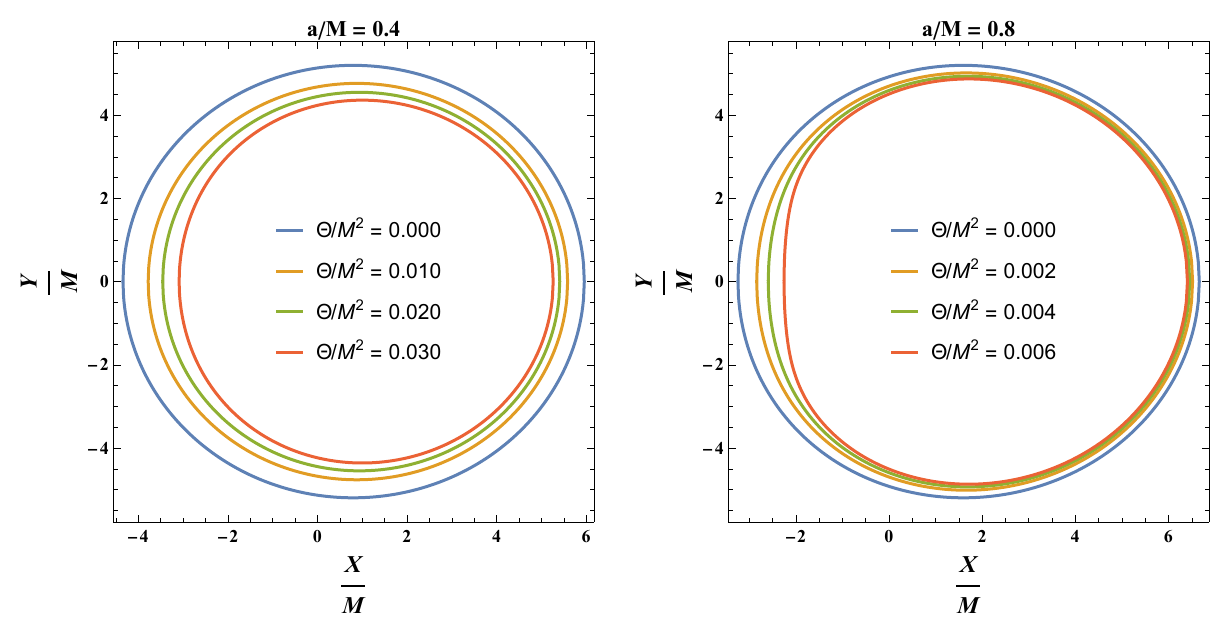}
		\caption{Shadow boundaries for an equatorial observer. The deformation changes both the overall angular scale and, for larger spin, the horizontal asymmetry. Only parameter values inside the horizon-existence region are shown.}
		\label{fig:shadows}
	\end{figure}

	For an observer at infinity at inclination $\theta_{o}$, the celestial coordinates are
	\begin{equation}
		X=-\xi_{p}\csc\theta_{o},
		\qquad
		Y=\pm\sqrt{\eta_{p}+a^{2}\cos^{2}\theta_{o}
			-\xi_{p}^{2}\cot^{2}\theta_{o}}.
		\label{eq:celestial}
	\end{equation}
	For an equatorial observer,
	\begin{equation}
		X=-\xi_{p},
		\qquad Y=\pm\sqrt{\eta_{p}}.
	\end{equation}
	The shadow boundary is generated by varying $r_p$ over the unstable
	spherical-photon family satisfying the angular admissibility condition
	given in Eq.~(114).
	 The corresponding shadow boundaries
	are displayed in  FIG.~\ref{fig:shadows}.

	\section{Timelike equatorial circular motion}
	\label{sec:timelike}
	
	Timelike equatorial circular geodesics determine the orbital
	energetics of matter in the vicinity of the black hole and provide a
	first approximation to the inner regions of a thin accretion disk.
	Their existence and stability are controlled by the radial geodesic
	potential. In this section we derive the specific energy, angular
	momentum, and orbital frequency of the prograde and retrograde
	branches and then determine how the non-commutative deformation shifts
	the marginally stable orbit. Particular attention is paid to the ISCO,
	which sets the conventional inner edge of a geodesic thin disk.

	\subsection{Conserved quantities and radial equation}
	
	For a massive particle with four-velocity $U^{\mu}$,
	\begin{equation}
		E=-U_{t}=-g_{tt}U^{t}-g_{t\phi}U^{\phi},
		\qquad
		L=U_{\phi}=g_{t\phi}U^{t}+g_{\phi\phi}U^{\phi}.
	\end{equation}
	Solving it gives
	\begin{equation}
		U^{t}=\frac{Eg_{\phi\phi}+Lg_{t\phi}}
			{g_{t\phi}^{2}-g_{tt}g_{\phi\phi}},
		\qquad
		U^{\phi}=-\frac{Eg_{t\phi}+Lg_{tt}}
			{g_{t\phi}^{2}-g_{tt}g_{\phi\phi}}.
	\end{equation}
	The normalization $U^{\mu}U_{\mu}=-1$ becomes
	\begin{equation}
		g_{rr}(U^{r})^{2}+g_{\theta\theta}(U^{\theta})^{2}+1
		-\frac{E^{2}g_{\phi\phi}+2ELg_{t\phi}+L^{2}g_{tt}}
		{g_{t\phi}^{2}-g_{tt}g_{\phi\phi}}=0.
		\label{eq:normalization-correct}
	\end{equation}
	For equatorial motion, $U^{\theta}=0$, and the radial polynomial can be written as
	\begin{equation}
		r^{4}(U^{r})^{2}=\cR_{m}(r),
	\end{equation}
	where
	\begin{equation}
		\cR_{m}(r)=\left[E(r^{2}+a^{2})-aL\right]^{2}
		-\Delta\left[r^{2}+(L-aE)^{2}\right].
		\label{eq:massive-radial}
	\end{equation}
	This radial polynomial should not be identified directly with the dimensionless effective potential in Eq.~\eqref{eq:normalization-correct}; they differ by metric-dependent factors.

	\subsection{Energy, angular momentum, and angular velocity}
	
	Let us define
	\begin{equation}
		A(r)=\sqrt{Mr-\beff},
		\qquad
		D_{s}(r)=r^{2}-3Mr+2\beff+2s aA(r),
		\qquad s=\pm1.
		\label{eq:As-Ds}
	\end{equation}
	The sign $s$ labels the orbit of the \emph{massive emitter}: for $a>0$, $s=+1$ is the prograde branch and $s=-1$ the retrograde branch. Circular equatorial motion satisfies $\cR_m=\cR_m'=0$, yielding
	\begin{equation}
		E_{s}=
		\frac{r^{2}-2Mr+\beff+s aA(r)}
		{r\sqrt{D_{s}(r)}},
		\qquad
		L_s=
		\frac{
			s\left\{
			A(r)\left[r^2+a^2-2saA(r)\right]
			-sa\beta_\Theta
			\right\}}
		{r\sqrt{D_s(r)}}.
	\end{equation}
	The corresponding four-velocity components are
	\begin{equation}
		U^{t}_{s}=
		\frac{r^{2}+s aA(r)}{r\sqrt{D_{s}(r)}},
		\qquad
		U^{\phi}_{s}=
		\frac{s A(r)}{r\sqrt{D_{s}(r)}},
		\label{eq:Uphis}
	\end{equation}
	and the orbital angular velocity is
	\begin{equation}
		\Omega_{s}=\frac{U^{\phi}_{s}}{U^{t}_{s}}
			=\frac{s A(r)}{r^{2}+s aA(r)}.
		\label{eq:orbitalOmega}
	\end{equation}
	The condition $D_s\geq0$ defines the circular-orbit domain, while $D_s=0$ reproduces the corresponding equatorial photon-radius equation.

	\subsection{Marginally bound orbit and ISCO}
	
	The marginally bound radius $r_{\rm mb}^{s}$ is defined by
	\begin{equation}
		E_{s}(r_{\rm mb}^{s})=1.
	\end{equation}
	In the Kerr limit,
	\begin{equation}
		r_{\rm mb}^{s}=2M-s a+2\sqrt{M(M-s a)}.
	\end{equation}
	The ISCO follows from $\cR_m=\cR_m'=\cR_m''=0$. An equivalent equation is
	\begin{align}
		0={}&Mr^{2}(r-6M)+4\beff(a^{2}-\beff)
		+3Mr(3\beff-a^{2})
		+8s a(Mr-\beff)^{3/2}.
		\label{eq:isco}
	\end{align}
	The stable branch satisfies the left-hand side of Eq.~\eqref{eq:isco} greater than or equal to zero, with equality at $r=r_{\rm ISCO}^{s}$.
	
	To display the effect of non-commutativity, introduce
	\begin{equation}
		x=\frac{r}{M},
		\qquad
		\alpha=\frac{a}{M},
		\qquad
		q=\frac{\beff}{M^{2}}
		=\frac{8\sqrt{\Theta}}{\sqrt{\pi}M}.
		\label{eq:isco-dimensionless}
	\end{equation}
	Then the ISCO equation becomes
	\begin{equation}
		x^{2}(x-6)+3x(3q-\alpha^{2})
			+4q(\alpha^{2}-q)+8s\alpha(x-q)^{3/2}=0.
		\label{eq:isco-dimensionless-equation}
	\end{equation}
	For a non-rotating black hole, $\alpha=0$, this reduces to
	\begin{equation}
		x^{3}-6x^{2}+9qx-4q^{2}=0.
	\end{equation}
	The outer physical root has the small-$q$ expansion
	\begin{equation}
		x_{\rm ISCO}=6-\frac{3}{2}q-\frac{19}{72}q^{2}
		+\mathcal{O}(q^{3}),
	\end{equation}
	so, through second order in $\sqrt{\Theta}$,
	\begin{equation}
		r_{\rm ISCO}=6M
			-\frac{12}{\sqrt{\pi}}\sqrt{\Theta}
			-\frac{152}{9\pi}\frac{\Theta}{M}
			+\mathcal{O}\!\left(\frac{\Theta^{3/2}}{M^{2}}\right).
		\label{eq:isco-static-expansion}
	\end{equation}
	Both corrections are negative: at fixed $M$, a nonzero non-commutative  parameter moves the static ISCO inside $6M$. At the extremal static endpoint $q=1$, the polynomial factorizes as $(x-1)^{2}(x-4)=0$, and the exterior ISCO is $r_{\rm ISCO}=4M$.
	
	For rotating configurations, let $x_{\rm Kerr}^{s}$ be the Kerr ISCO root and write
	\begin{equation}
		x_{\rm ISCO}^{s}=x_{\rm Kerr}^{s}+c_{s}q+\mathcal{O}(q^{2}).
	\end{equation}
	The linear coefficient is
	\begin{equation}
		c_{s}=-\frac{9x_{\rm Kerr}^{s}+4\alpha^{2}
				-12s\alpha\sqrt{x_{\rm Kerr}^{s}}}
			{3(x_{\rm Kerr}^{s})^{2}-12x_{\rm Kerr}^{s}
				-3\alpha^{2}+12s\alpha\sqrt{x_{\rm Kerr}^{s}}}.
		\label{eq:isco-cs}
	\end{equation}
	For the physical nonextremal prograde and retrograde branches in the parameter range considered here, $c_s<0$; hence increasing $\Theta$ shifts both ISCOs inward. The comparison with $6M$ is therefore:
	\begin{equation}
		\begin{array}{ll}
			a=0,\ \Theta=0: & r_{\rm ISCO}=6M,\\ [1ex]
			a=0,\ \Theta>0: & r_{\rm ISCO}<6M,\\ [1.5ex]
			a>0,\ s=+1: & r_{\rm ISCO}^{\rm pro}<6M\ 
			\text{and decreases further with }\Theta,\\ [1.5ex]
			a>0,\ s=-1: & r_{\rm ISCO}^{\rm retro}>6M\ \text{at }\Theta=0,\ 
			\text{but decreases as }\Theta\text{ grows.}
		\end{array}
		\label{eq:isco-summary}
	\end{equation}   
	The retrograde branch may remain above $6M$ for weak deformation, but it can cross below $6M$ when the non-commutative  correction is sufficiently strong. For example, at $M=1$, $a=0.1$, and $\Theta=0.01$,
	\begin{equation}
		r_{\rm ISCO}^{\rm pro}\simeq4.897M,
		\qquad
		r_{\rm ISCO}^{\rm retro}\simeq5.604M,
	\end{equation}
	whereas the corresponding Kerr values are approximately $5.669M$ and $6.323M$. The inward shift is analogous to the charge-induced displacement in Kerr--Newman geometry, with the important difference that $\beff$ is constrained by $M$ and $\Theta$.

	\section{Photon emission and frequency shifts}
	\label{sec:redshift}
	
The frequency measured by an observer is the invariant scalar
$\omega=-k_{\mu}U^{\mu}$. Consequently, the observed shift combines
the gravitational contribution, the special-relativistic Doppler
effect, and rotational frame dragging in a coordinate-independent
manner. We consider an emitter following a circular equatorial
timelike geodesic and a photon emitted tangentially in the local
equatorial plane. The emitter-orbit sign and photon-direction sign
must be kept independent, and a photon assigned to a distant detector
must additionally satisfy the global escape condition determined by
its radial potential.

	\subsection{Emitter orbit and photon direction}
	
	The emitter is a massive source, for example a gas element or atom in an accretion flow. For a circular equatorial orbit,
	\begin{equation}
		U_{e,s}^{\mu}=(U_{e,s}^{t},0,0,U_{e,s}^{\phi}),
		\qquad r=r_e,\qquad \theta=\frac{\pi}{2},
		\qquad s=\pm1,
		\label{eq:emitter-vector-s}
	\end{equation}
	where $s=+1$ and $s=-1$ denote prograde and retrograde emitter orbits, respectively. The components are obtained from Eqs.~\eqref{eq:Uphis} evaluated at $r_e$:
	\begin{equation}
		U_{e,s}^{t}=\frac{r_e^{2}+s aA_e}{r_e\sqrt{D_{e,s}}},\qquad
		U_{e,s}^{\phi}=\frac{sA_e}{r_e\sqrt{D_{e,s}}},
		\label{eq:emitter-components}
	\end{equation}
	with
	\begin{equation}
		A_e=\sqrt{Mr_e-\beff},
		\qquad
		D_{e,s}=r_e^{2}-3Mr_e+2\beff+2s aA_e.
	\end{equation}
	The statement $U_{e,s}^{r}=0$ means that the emitter has no radial velocity; it does not imply that the emitted photon has $k^{r}=0$. In the emitter's local rest frame,
	\begin{equation}
		k^{\mu}=\omega_e(U_{e,s}^{\mu}+n^{\mu}),
		\qquad U_{e,s\,\mu}n^{\mu}=0,
		\qquad n_{\mu}n^{\mu}=1,
	\end{equation}
	so the photon can be emitted outward, inward, or tangentially according to the spatial direction $n^{\mu}$.

	\subsection{Photon impact parameter and the sign $\sigma$}
	
	The photon satisfies
	$k^{\mu}k_{\mu}=0$,
	and possesses the conserved quantities
	\begin{equation}
		E_{\gamma}=-k_t,
		\qquad
		L_{\gamma}=k_{\phi},
		\qquad
		b=\frac{L_{\gamma}}{E_{\gamma}}.
	\end{equation}
	For instantaneous tangential emission, $k^{r}=k^{\theta}=0$, the null condition becomes
	\begin{equation}
		g_{tt}b^{2}+2g_{t\phi}b+g_{\phi\phi}=0.
	\end{equation}
	We label the two photon directions by a second, independent sign $\sigma=\pm1$ and define
	\begin{equation}
		b_{\sigma}=
			\frac{-g_{t\phi}-\sigma\sqrt{g_{t\phi}^{2}-g_{tt}g_{\phi\phi}}}
			{g_{tt}},
		\qquad \sigma=\pm1.
		\label{eq:bsigma-general}
	\end{equation}
	With this convention, sufficiently far from a black hole with $a>0$, $\sigma=+1$ gives $b_{\sigma}>0$ and $\sigma=-1$ gives $b_{\sigma}<0$. On the equatorial plane,
	\begin{equation}
		b_{\sigma}(r)=
			\frac{-a(2Mr-\beff)+\sigma r^{2}\sqrt{\Delta}}
			{r^{2}-2Mr+\beff}.
		\label{eq:bsigma-explicit}
	\end{equation}
	The sign $\sigma$ labels the photon emission direction, whereas $s$ labels the emitter orbit. They are independent choices. In particular, a prograde emitter $(s=+1)$ can emit either a $\sigma=+1$ or a $\sigma=-1$ photon. The algebraic label $\sigma$ should also not be identified automatically with redshift or blueshift; that classification follows from the final value of $z$.

	\subsection{Invariant frequency shifts with $(s,\sigma)$ notation}
	
	An observer with four-velocity $U^{\mu}$ measures
	$\omega=-k_{\mu}U^{\mu}$.
	For circular equatorial emission and detection,
	\begin{equation}
		\omega=E_{\gamma}(U^{t}-b_{\sigma}U^{\phi}).
	\end{equation}
	The total frequency shift is therefore
	\begin{equation}
		1+z_{\sigma;s}=
			\frac{U_{e,s}^{t}-b_{\sigma}U_{e,s}^{\phi}}
			{U_{d}^{t}-b_{\sigma}U_{d}^{\phi}}.
		\label{eq:zsigmas-general}
	\end{equation}
	The semicolon emphasizes that $\sigma$ and $s$ refer to different objects. There are four possible combinations,
	\begin{equation}
		(s,\sigma)=(+,+),(+,-),(-,+),(-,-).
	\end{equation}
	For each fixed emitter branch $s$, the same $U_{e,s}^{t}$ and $U_{e,s}^{\phi}$ must be used for both photon branches $b_{+}$ and $b_{-}$.
	
	For a distant static observer,
	\begin{equation}
		U_d^{t}\rightarrow1,
		\qquad U_d^{\phi}\rightarrow0,
	\end{equation}
	and Eq.~\eqref{eq:zsigmas-general} reduces to
	\begin{equation}
		1+z_{\sigma;s}
			=U_{e,s}^{t}-b_{\sigma}(r_e)U_{e,s}^{\phi}
		\label{eq:zsigmas-distant}
	\end{equation}
	or explicitly,
	\begin{equation}
		z_{\sigma;s}=
			\frac{r_e^{2}+sA_e[a-b_{\sigma}(r_e)]}
			{r_e\sqrt{D_{e,s}}}-1.
		\label{eq:zsigmas-explicit}
	\end{equation}
	The emitter radius must lie in the physical timelike circular-orbit domain, and for a stable disk emitter one should choose
	$r_e\geq r_{\rm ISCO}^{s}$.
	A natural inner-edge choice is $r_e=r_{\rm ISCO}^{s}$, while a fixed larger radius may be used for controlled comparisons; in either case, $r_e$ must be stated in every plot.
	
\begin{figure}[htb]
		\centering
		\safeincludegraphics[width=.45\linewidth]{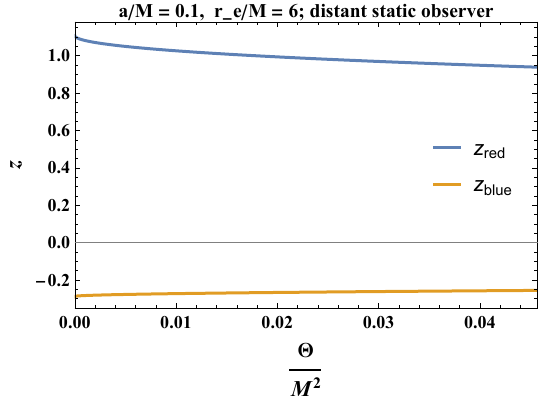}
		\caption{Frequency shifts for a prograde emitter $(s=+1)$ at $r_e/M=6$ and a distant static observer. The two curves correspond to the two tangential photon branches $\sigma=\pm1$; they are classified as redshift or blueshift only after evaluating $z_{\sigma;+}$.}
		\label{fig:frequency-shifts}
	\end{figure}

	With the convention
	\begin{equation}
		1+z=\frac{\omega_e}{\omega_d},
	\end{equation}
	redshift corresponds to $z_{\sigma;s}>0$, while blueshift corresponds to $-1<z_{\sigma;s}<0$. Thus neither $s$ nor $\sigma$ alone determines whether the received photon is redshifted or blueshifted.
	  The frequency redshift or blueshift of a prograde emitter are shown in FIG.~\ref{fig:frequency-shifts}.
	 
	\section{Discussion and conclusions}
	\label{sec:conclusions}
	
	We have presented a corrected and extended analysis of an effective Kerr black hole in a non-commutative  background. The deformation appears in the radial function through $\beff=8M\sqrt{\Theta/\pi}$. Algebraically, the metric belongs to the Kerr--Newman family under the formal replacement $Q^{2}\to\beff$, but the physical interpretation is different because $\beff$ is not an independent charge.
	
	The horizon equation yields a sharp upper bound on $\Theta$. At saturation, the outer and inner horizons merge, the Hawking temperature vanishes, and the black hole reaches a remnant with $M_{\rm rem}=r_{\rm rem}$. The stationary-limit surfaces were obtained separately from the horizons, and their difference defines the ergoregion. The stationary-limit surface touches the horizon at the poles, reaches its largest separation at the equator, disappears as a distinct surface when $a=0$, and reduces to the familiar Kerr expression when $\Theta=0$.
	
	A central improvement is the separation between geometric and thermodynamic quantities. Since $\beff=M\lth$, the geometric pair $(T_H,\Omega_H)$ does not coincide with the exact conjugate pair $(\widetilde T,\widetilde\Omega)$ of the constrained fixed-$\Theta$ fundamental relation. We therefore treated the strict fixed-$J$ canonical ensemble using $\widetilde F_J$ and $\widetilde C_{J,\Theta}$, and the strict grand-canonical ensemble using $\widetilde G$ and $\widetilde C_{\widetilde\Omega,\Theta}$. The response $C_{\Omega_H,\Theta}^{(H)}$ was retained only as a Hawking-temperature response along fixed geometric $\Omega_H$. These critical points describe different paths through state space and must not be mixed. Any future thermodynamic-topology analysis should be constructed from the potential appropriate to the chosen ensemble.
	
	For null motion, the rotating geometry possesses a photon region rather than a single photon sphere. Its equatorial endpoints are the prograde and retrograde light rings, while intermediate radii correspond to non-equatorial spherical photon trajectories. The shadow follows from the same impact parameters. For massive circular motion, the non-commutative  parameter moves the physical ISCO branches inward. In the static case the small-$\Theta$ expansion begins at $6M$ and receives negative $\sqrt{\Theta}$ and $\Theta$ corrections, approaching $4M$ at the extremal static endpoint. For rotating configurations, both prograde and retrograde ISCOs decrease over the physical parameter range examined; the retrograde branch can therefore cross below $6M$ for sufficiently strong deformation.
	
	Finally, the frequency-shift calculation was reformulated with two independent signs. The emitter sign $s=\pm1$ specifies the prograde or retrograde massive circular orbit and determines $U_{e,s}^{t}$ and $U_{e,s}^{\phi}$. The photon sign $\sigma=\pm1$ specifies the tangential null direction and determines $b_{\sigma}$. For each fixed emitter branch, the same emitter four-velocity must be used for both photon directions. The observable $z_{\sigma;s}$, rather than either sign alone, decides whether the received photon is redshifted or blueshifted.
	
	The present model retains the Kerr ring singularity at $\Sigma=0$ and should therefore be described as an effective non-commutative  deformation rather than a fully regular rotating black hole. Future work may address a rotating geometry derived from a genuinely smeared mass function, a full extended first law with $\Theta$ varied as a thermodynamic variable, ensemble-dependent thermodynamic topology, ray-traced images at finite observer distance, and accretion-disk spectra computed from the corrected circular-orbit structure.

	\section*{Acknowledgments}
	H.H. is grateful to Excellence project FoS UHK 2203/2025-2026 for the financial support.
	The research of L.M.N. was supported by PID2023-148409NB-I00, funded by MICIU/AEI/10.13039/501100011033, and by  the Department of Education of the Junta de Castilla y Leon and FEDER Funds (Reference: CLU-2023-1-05).
N.B. Acknowledges University of Valladolid for hospitality during part of this work.


\begin{thebibliography}{99}
		
		\bibitem{SeibergWitten}
		N.~Seiberg and E.~Witten,
		``String theory and noncommutative geometry,''
		JHEP \textbf{09} (1999) 032.
		
		\bibitem{Szabo}
		R.~J.~Szabo,
		``Quantum field theory on noncommutative spaces,''
		Phys. Rept. \textbf{378} (2003) 207--299.
		
		\bibitem{DouglasNekrasov}
		M.~R.~Douglas and N.~A.~Nekrasov,
		``Noncommutative field theory,''
		Rev. Mod. Phys. \textbf{73} (2001) 977--1029.
		
		\bibitem{NicoliniReview}
		P.~Nicolini,
		``Noncommutative black holes, the final appeal to quantum gravity: a review,''
		Int. J. Mod. Phys. A \textbf{24} (2009) 1229--1308.
		
		\bibitem{SmailagicSpallucci}
		A.~Smailagic and E.~Spallucci,
		``Feynman path integral on the noncommutative plane,''
		J. Phys. A \textbf{36} (2003) L467--L471.
		
		\bibitem{Myung}
		Y.~S.~Myung, Y.-W.~Kim, and Y.-J.~Park,
		``Thermodynamics and evaporation of the noncommutative black hole,''
		JHEP \textbf{02} (2007) 012.
		
		\bibitem{ModestoNicolini}
		L.~Modesto and P.~Nicolini,
		``Charged rotating noncommutative black holes,''
		Phys. Rev. D \textbf{82} (2010) 104035.
		
		\bibitem{banerjee2008noncommutative}
		R. Banerjee, B.R. Majhi, and S. Samanta, ``Noncommutative black hole thermodynamics,'' Physical Review D—Particles, Fields, Gravitation, and Cosmology, 77(12) (2008) 124035.

		\bibitem{nozari2006reissner}
		K. Nozari and B. Fazlpour, arXiv:gr-qc/0608077.
		
		\bibitem{sharif2011thermodynamics}
		M. Sharif and W. Javed, ``Thermodynamics of a Bardeen black hole in noncommutative space,'' Canadian Journal of Physics, 89(10) (2011) 1027-1033.
		
		\bibitem{heidari2025non}
		N. Heidari, A.A. Araújo Filho and I.P. Lobo, ``Non-commutativity in Hayward spacetime,'' JCAP, \textbf{09} (2025) 051.
		
		\bibitem{araujo2025non}
		A.A. Araújo Filho, N. Heidari, and I.P. Lobo, ``A non-commutative Kalb-Ramond black hole,'' JCAP, \textbf{09} (2025) 076.

		\bibitem{zhao2024quasinormal}
		Y. Zhao, Y. Cai, S. Das, G. Lambiase, E.N. Saridakis and E.C. Vagenas,
		``Quasinormal modes in noncommutative Schwarzschild black holes,'' Nuclear Physics B, 1004 (2024) 116545.
		
		\bibitem{anacleto2023absorption}
		M.A. Anacleto, F.A. Brito, J.A.V. Campos and E. Passos, ``Absorption, scattering and shadow by a noncommutative black hole with global monopole,'' Eur. Phys. J. C, 83(4) (2023) 298.

		\bibitem{anacleto2020absorption}
		M.A. Anacleto, F.A. Brito, J.A.V. Campos and E. Passos,
		``Absorption and scattering of a noncommutative black hole,''
		Phys.\ Lett.\ B \textbf{803}, 135334 (2020).

		\bibitem{heidari2024exploring}
		N. Heidari, H. Hassanabadi, A.A. Araújo Filho and J. Kriz, ``Exploring non-commutativity as a perturbation in the Schwarzschild black hole: quasinormal modes, scattering, and shadows,'' Eur. Phys. J. C, 84(6) (2024) 566.
		
		\bibitem{chaichian2008corrections}
		M. Chaichian, A. Tureanu and G. Zet,
		``Corrections to Schwarzschild solution in noncommutative gauge theory of gravity,''
		Phys.\ Lett.\ B \textbf{660}, 573--578 (2008).		
		
		\bibitem{Kerr}
		R.~P.~Kerr,
		``Gravitational field of a spinning mass as an example of algebraically special metrics,''
		Phys. Rev. Lett. \textbf{11} (1963) 237--238.
		
		\bibitem{Carter}
		B.~Carter,
		``Global structure of the Kerr family of gravitational fields,''
		Phys. Rev. \textbf{174} (1968) 1559--1571.
		
		\bibitem{Chandrasekhar}
		S.~Chandrasekhar,
		\emph{The Mathematical Theory of Black Holes},
		Oxford University Press, Oxford (1983).
		
		\bibitem{Bardeen}
		J.~M.~Bardeen,
		``Timelike and null geodesics in the Kerr metric,''
		in \emph{Black Holes}, eds. C.~DeWitt and B.~S.~DeWitt,
		Gordon and Breach, New York (1973), pp. 215--239.
		
		\bibitem{Teo}
		E.~Teo,
		``Spherical photon orbits around a Kerr black hole,''
		Gen. Rel. Grav. \textbf{35} (2003) 1909--1926.
		
	
		
		\bibitem{Bardeen1972}
		J.~M.~Bardeen, W.~H.~Press, and S.~A.~Teukolsky,
		``Rotating black holes: Locally nonrotating frames, energy extraction,
		and scalar synchrotron radiation,''
		Astrophys.\ J.\ \textbf{178}, 347--370 (1972).
		
			\bibitem{Pugliese2011}
		D.~Pugliese, H.~Quevedo, and R.~Ruffini,
		``Equatorial circular motion in Kerr spacetime,''
		Phys.\ Rev.\ D \textbf{84}, 044030 (2011).
		
		
		\bibitem{NovikovThorne1973}
		I.~D.~Novikov and K.~S.~Thorne,
		``Astrophysics of black holes,''
		in \textit{Black Holes (Les Astres Occlus)},
		edited by C.~DeWitt and B.~S.~DeWitt
		(Gordon and Breach, New York, 1973), pp.~343--450.
		
		\bibitem{PageThorne1974}
		D.~N.~Page and K.~S.~Thorne,
		``Disk-accretion onto a black hole. I. Time-averaged structure of
		accretion disk,''
		Astrophys.\ J.\ \textbf{191}, 499--506 (1974).
		
		\bibitem{Thorne1974}
		K.~S.~Thorne,
		``Disk-accretion onto a black hole. II. Evolution of the hole,''
		Astrophys.\ J.\ \textbf{191}, 507--520 (1974).
		
		\bibitem{Cunningham1975}
		C.~T.~Cunningham,
		``The effects of redshifts and focusing on the spectrum of an accretion
		disk around a Kerr black hole,''
		Astrophys.\ J.\ \textbf{202}, 788--802 (1975).
		
	
		
		\bibitem{StuchlikSlany2004}
		Z.~Stuchl\'{\i}k and P.~Slan\'y,
		``Equatorial circular orbits in the Kerr--de Sitter spacetimes,''
		Phys.\ Rev.\ D \textbf{69}, 064001 (2004).
		
		\bibitem{JohannsenPsaltis2011}
		T.~Johannsen and D.~Psaltis,
		``A metric for rapidly spinning black holes suitable for strong-field
		tests of the no-hair theorem,''
		Phys.\ Rev.\ D \textbf{83}, 124015 (2011).
		
		\bibitem{Bambi2017}
		C.~Bambi,
		``Testing black hole candidates with electromagnetic radiation,''
		Rev.\ Mod.\ Phys.\ \textbf{89}, 025001 (2017).
		
		\bibitem{StuchlikSchee2015}
		Z.~Stuchl\'{\i}k and J.~Schee,
		``Circular geodesics of Bardeen and Ay\'on-Beato--Garc\'{\i}a regular
		black-hole and no-horizon spacetimes,''
		Int.\ J.\ Mod.\ Phys.\ D \textbf{24}, 1550020 (2015).
		
	\bibitem{Nicolini2006}
		P.~Nicolini, A.~Smailagic, and E.~Spallucci,
		``Noncommutative geometry inspired Schwarzschild black hole,''
		Phys.\ Lett.\ B \textbf{632}, 547--551 (2006).
		
		\bibitem{Ansoldi2007}
		S.~Ansoldi, P.~Nicolini, A.~Smailagic, and E.~Spallucci,
		``Noncommutative geometry inspired charged black holes,''
		Phys.\ Lett.\ B \textbf{645}, 261--266 (2007).
		
		
		\bibitem{NicoliniSpallucci2010}
		P.~Nicolini and E.~Spallucci,
		``Noncommutative geometry inspired dirty black holes,''
		Class.\ Quantum Grav.\ \textbf{27}, 015010 (2010).
		
		\bibitem{SmailagicSpallucci2010}
		A.~Smailagic and E.~Spallucci,
		```Kerrr' black hole: The Lord of the String,''
		Phys.\ Lett.\ B \textbf{688}, 82--87 (2010).
		
		\bibitem{ModestoNicolini2010}
		L.~Modesto and P.~Nicolini,
		``Charged rotating noncommutative black holes,''
		Phys.\ Rev.\ D \textbf{82}, 104035 (2010).
		
		\bibitem{Wei2015}
		S.-W.~Wei, P.~Cheng, Y.~Zhong, and X.-N.~Zhou,
		``Shadow of noncommutative geometry inspired black hole,''
		JCAP \textbf{08}, 004 (2015).
		
		\bibitem{KumarGhosh2017}
		R.~Kumar and S.~G.~Ghosh,
		``Accretion onto a noncommutative geometry inspired black hole,''
		Eur.\ Phys.\ J.\ C \textbf{77}, 577 (2017).
		
	
		\bibitem{Anacleto2020}
		M.~A.~Anacleto, F.~A.~Brito, J.~A.~V.~Campos, and E.~Passos,
		``Absorption and scattering of a noncommutative black hole,''
		Phys. Lett. B \textbf{803} (2020) 135334.
		
		\bibitem{araujo2024effects}
		Ara{\'u}jo Filho, A. A., et al. "Effects of non-commutative geometry on black hole properties." Physics of the Dark Universe 46 (2024): 101630.

\bibitem{CapozzielloJCAP2023}
S Capozziello, S Zare, DF Mota, and  H Hassanabadi, JCAP 05 (2023) 027.
\bibitem{CapozzielloPoDU2025}
S. Capozziello, S. Zare, L. M. Nieto, and  H. Hassanabadi, Phys. Dark Universe 50 (2025) 102065.

		

\bibitem{NewmanJanis1965}
E.~T.~Newman and A.~I.~Janis,
``Note on the Kerr spinning-particle metric,''
J.\ Math.\ Phys.\ \textbf{6}, 915--917 (1965).
\href{https://doi.org/10.1063/1.1704350}
{doi:10.1063/1.1704350}.
		
		\bibitem{NewmanEtAl1965}
		E.~T.~Newman, E.~Couch, K.~Chinnapared, A.~Exton, A.~Prakash,
		and R.~Torrence,
		``Metric of a rotating, charged mass,''
		J.\ Math.\ Phys.\ \textbf{6}, 918--919 (1965).
		\href{https://doi.org/10.1063/1.1704351}
		{doi:10.1063/1.1704351}.
		
		\bibitem{DrakeSzekeres2000}
		S.~P.~Drake and P.~Szekeres,
		``Uniqueness of the Newman--Janis algorithm in generating the
		Kerr--Newman metric,''
		Gen.\ Relativ.\ Gravit.\ \textbf{32}, 445--458 (2000).
		\href{https://doi.org/10.1023/A:1001920232180}
		{doi:10.1023/A:1001920232180}.
		
		\bibitem{BambiModesto2013}
		C.~Bambi and L.~Modesto,
		``Rotating regular black holes,''
		Phys.\ Lett.\ B \textbf{721}, 329--334 (2013).
		\href{https://doi.org/10.1016/j.physletb.2013.03.025}
		{doi:10.1016/j.physletb.2013.03.025}.
		

		
		\bibitem{AzregAinou2014}
		M.~Azreg-A\"{\i}nou,
		``Generating rotating regular black hole solutions without
		complexification,''
		Phys.\ Rev.\ D \textbf{90}, 064041 (2014).
		\href{https://doi.org/10.1103/PhysRevD.90.064041}
		{doi:10.1103/PhysRevD.90.064041}.
		
		\bibitem{AzregAinou2014EPJC}
		M.~Azreg-A\"{\i}nou,
		``From static to rotating to conformal static solutions: Rotating
		imperfect fluid wormholes with(out) electric or magnetic field,''
		Eur.\ Phys.\ J.\ C \textbf{74}, 2865 (2014).
		\href{https://doi.org/10.1140/epjc/s10052-014-2865-8}
		{doi:10.1140/epjc/s10052-014-2865-8}.
		
		\bibitem{Erbin2017}
		H.~Erbin,
		``Janis--Newman algorithm: Generating rotating and NUT charged
		black holes,''
		Universe \textbf{3}, 19 (2017).
		\href{https://doi.org/10.3390/universe3010019}
		{doi:10.3390/universe3010019}.
		
		\bibitem{Shaikh2019}
		R.~Shaikh,
		``Black hole shadow in a general rotating spacetime obtained through
		the Newman--Janis algorithm,''
		Phys.\ Rev.\ D \textbf{100}, 024028 (2019).
		\href{https://doi.org/10.1103/PhysRevD.100.024028}
		{doi:10.1103/PhysRevD.100.024028}.
		
		\bibitem{LimaJunior2020}
		H.~C.~D.~Lima Junior, L.~C.~B.~Crispino, P.~V.~P.~Cunha,
		and C.~A.~R.~Herdeiro,
		``Spinning black holes with a separable Hamilton--Jacobi equation
		from a modified Newman--Janis algorithm,''
		Eur.\ Phys.\ J.\ C \textbf{80}, 1036 (2020).
		\href{https://doi.org/10.1140/epjc/s10052-020-08572-w}
		{doi:10.1140/epjc/s10052-020-08572-w}.
		
				\bibitem{HansenYunes2013}
		D.~Hansen and N.~Yunes,
		``Applicability of the Newman--Janis algorithm to black hole solutions
		of modified gravity theories,''
		Phys.\ Rev.\ D \textbf{88}, 104020 (2013).
		\href{https://doi.org/10.1103/PhysRevD.88.104020}
		{doi:10.1103/PhysRevD.88.104020}.
				
		
\bibitem{PerlickTsupko}
V.~Perlick, and O.~Y.~Tsupko,
``Calculating black hole shadows: review of analytical studies,''
Phys. Rept. \textbf{947} (2022) 1--39.
		
		
\bibitem{ZarePLB2024}		
S. Zare, L. M. Nieto, F. Hosseinifar, X.-H. Feng, and H. Hassanabadi, Phys. Lett. B 859 (2024) 139125.
\bibitem{ZareJCAP2024}
S. Zare, L. M. Nieto, X.-H. Feng, S.-H. Dong, and H. Hassanabadi, JCAP 08 (2024) 041
\bibitem{ZareJCAP2026}
S. Zare, T. Zhu, L. M. Nieto, S. Lu, and H. Hassanabadi, JCAP 01 (2026) 059.
\bibitem{ZareEPJC2026}
S. Zare, F. Hosseinifar, L. M. Nieto, D. J. Gogoi, K. Boshkayev, A. Urazalina, and H. Hassanabadi, Eur. Phys. J. C 86 (2026) 160.
\bibitem{Hassanabadi2026-1}
H. Hassanabadi, A. Guvendi, F. Kafikang, T. Sathiyaraj, and S. Zare, arXiv:2512.18512
\bibitem{Hosseinifar2026}
F. Hosseinifar, S. Mamedov, K. Boshkayev, S. Zare, F. Studni\v{c}ka, and H. Hassanabadi, 	arXiv:2605.20239
\bibitem{Hassanabadi2026-2}
H. Hassanabadi, M. R. R. Good, S. Zare, O. Luongo and F. Kafikang, 	arXiv:2607.11979



		
	\bibitem{HerreraNucamendi}
		A.~Herrera-Aguilar and U.~Nucamendi,
		``Kerr black hole parameters in terms of the redshift/blueshift of photons emitted by geodesic particles,''
		Phys. Rev. D \textbf{92} (2015) 045024.
		
		\bibitem{Davies}
		P.~C.~W.~Davies,
		``Thermodynamics of black holes,''
		Rept. Prog. Phys. \textbf{41} (1978) 1313--1355.
		
		\bibitem{BanerjeeMajhi}
		R.~Banerjee, B.~R.~Majhi, and S.~Samanta,
		``Noncommutative black hole thermodynamics,''
		Phys. Rev. D \textbf{77} (2008) 124035.
		
		\bibitem{LopezDominguez}
		J.~C.~L\'opez-Dom\'inguez, O.~Obreg\'on, M.~Sabido, and C.~Ram\'irez,
		``Towards noncommutative quantum black holes,''
		Phys. Rev. D \textbf{74} (2006) 084024.

	\end{thebibliography}
\end{document}